\documentclass[a4paper]{article}

\usepackage{INTERSPEECH2018}
\usepackage{amsmath,amssymb,graphicx}
\usepackage{color}
\usepackage{balance}

\usepackage[]{algorithm2e}
\usepackage{wasysym}
\usepackage{enumitem}

\def\T{{\mathsf T}}

\def\F{{\mathrm F}}
\def\RR{{\mathbb R}}

\let\OLDthebibliography\thebibliography
\renewcommand\thebibliography[1]{
  \OLDthebibliography{#1}
  \setlength{\parskip}{0.5pt}
  \setlength{\itemsep}{0.5pt plus 0.3ex}
}

\title{End-to-End Speech Separation with Unfolded Iterative Phase Reconstruction}
\name{Zhong-Qiu Wang\thanks{\hspace{-.3cm}Part of this work was done while Z.-Q.\ Wang was an intern at MERL.}$^{1,2}$, Jonathan Le Roux$^{1}$, DeLiang Wang$^{2,3}$, John R. Hershey$^{1}$}
\address{$^{1}$Mitsubishi Electric Research Laboratories (MERL), USA \\
	$^{2}$Department of Computer Science and Engineering, The Ohio State University, USA \\
	$^{3}$Center for Cognitive and Brain Sciences, The Ohio State University, USA}
\email{\{wangzhon,dwang\}@cse.ohio-state.edu, leroux@merl.com}

\begin{document}

\maketitle
\setlength{\abovedisplayskip}{4pt}
\setlength{\belowdisplayskip}{4pt}
\begin{abstract}
This paper proposes an end-to-end approach for single-channel speaker-independent multi-speaker speech separation, where time-frequency (T-F) masking, the short-time Fourier transform (STFT), and its inverse are represented as  layers within a deep network. Previous approaches, rather than computing a loss on the reconstructed signal, used a surrogate loss based on the target STFT magnitudes. This ignores reconstruction error introduced by phase inconsistency. In our approach, the loss function is directly defined on the reconstructed signals, which are optimized for best separation. In addition, we train through unfolded iterations of a phase reconstruction algorithm, represented as a series of STFT and inverse STFT layers. While mask values are typically limited to lie between zero and one for approaches using the mixture phase for reconstruction, this limitation is less relevant if the estimated magnitudes are to be used together with phase reconstruction. We thus propose several novel activation functions for the output layer of the T-F masking, to allow mask values beyond one. On the publicly-available wsj0-2mix dataset, our approach achieves state-of-the-art 12.6~dB scale-invariant signal-to-distortion ratio (SI-SDR) and 13.1~dB SDR, revealing new possibilities for deep learning based phase reconstruction and representing a fundamental progress towards solving the notoriously-hard cocktail party problem.
\end{abstract}

\section{Introduction}
\label{sec:intro}

Recent years have witnessed exciting advances towards solving the cocktail party problem. The inventions of deep clustering \cite{Hershey2016, Isik2016, Wang2018ICASSP}, deep attractor networks \cite{Chen2017, Luo2018} and permutation free training \cite{Hershey2016, Isik2016, Yu2017a, Kolbæk2017} have dramatically improved the performance of single-channel speaker-independent multi-speaker speech separation, demonstrating overwhelming advantages over previous methods including graphical modeling approaches \cite{Hershey2010}, spectral clustering approaches \cite{Bach2006}, and CASA methods \cite{Wang2006}. 

However, all of these  conduct separation on the magnitude in the time-frequency (T-F) domain and directly use the mixture phase for time-domain re-synthesis, largely because phase is difficult to estimate. It is well-known that this incurs a phase inconsistency problem \cite{LeRoux2008SAPA09, Sturmel2011, Gerkmann2015}, especially for speech processing, where there is typically at least half overlap between consecutive frames. This overlap makes the STFT representation of a speech signal highly redundant. As a result, the enhanced STFT representation obtained using the estimated magnitude and mixture phase would not be in the consistent STFT domain, meaning that it is not guaranteed that there exists a time-domain signal having that STFT representation.

To improve the consistency, one stream of research is focused on iterative methods such as the classic Griffin-Lim algorithm \cite{Griffin1984a}, multiple input spectrogram inverse (MISI) \cite{Gunawan2010}, ISSIR \cite{Sturmel2013aslp}, and consistent Wiener filtering \cite{LeRoux2013SPL03}, which can recover the clean phase to some extent starting from the mixture phase and a good estimated magnitude by iteratively performing STFT and iSTFT \cite{Gerkmann2015}. There are some previous attempts at naively applying such iterative algorithms as a post-processing step on the magnitudes produced by deep learning based speech enhancement and separation \cite{Han2015, Zhao2017b, Li2016, Wang2018ICASSP}. However, this usually only leads to small improvements, even though the magnitude estimates from DNNs are reasonably good. We think that this is possibly because the T-F masking is performed without being aware of the later phase reconstruction steps and hence may not produce spectral structures that are appropriate for iterative phase reconstruction. 

This study hence proposes a novel end-to-end speech separation algorithm that trains through iterative phase reconstruction via T-F masking for signal-level approximation. On the publicly-available wsj0-2mix corpus, our algorithm reaches 12.6 dB scale-invariant SDR, which surpasses the previous best by a large margin and is comparable to the oracle 12.7 dB result obtained using the so-called ideal ratio mask (IRM). Our study shows, for the first time and based on a large open dataset, that deep learning based phase reconstruction leads to tangible and large improvements when combined with state-of-the-art magnitude-domain separation. 

\section{Chimera++ Network}

To elicit a good phase via phase reconstruction, it is necessary to first obtain a good enough magnitude estimate. Our recent study \cite{Wang2018ICASSP} proposed a novel multi-task learning approach
combining the regularization capability of deep clustering with the ease of end-to-end training of mask inference, yielding significant improvements over the individual models. 

The key idea of deep clustering \cite{Hershey2016} is to learn a high-dimensional embedding vector for each T-F unit using a powerful deep neural network (DNN) such that the embeddings of the T-F units dominated by the same speaker are close to each other in the embedding space while farther otherwise. This way, simple clustering methods like k-means can be applied to the learned embeddings to perform separation at run time. More specifically, the network computes a unit-length embedding vector ${v}_{i}\in \RR^{1\times D}$ corresponding to the $i^{th}$ T-F element. Similarly, $y_{i}\in \RR^{1\times C}$ is a one-hot label vector representing which source in a mixture dominates the $i^{th}$ T-F unit. Vertically stacking these, we form the embedding matrix $V\in \RR^{TF\times D}$ and the label matrix $Y\in \RR^{TF\times C}$. The embeddings are learned by approximating the affinity matrix from the embeddings:
\begin{align} \label{classicdpcl}
	\mathcal{L}_{\text{DC},\text{classic}} &= \|VV^{\T} - YY^{\T}\|_{\F}^2 
\end{align}

Our recent study \cite{Wang2018ICASSP} suggests that an alternative loss function, which whitens the embedding in a k-means objective, leads to better separation performance.
\begin{align}
    \mathcal{L}_{\text{DC},\text{W}} &=\|V(V^{\T}V)^{-\frac{1}{2}} -  Y(Y^{\T}Y)^{-1}Y^{\T} V (V^{\T}V)^{-\frac{1}{2}} \|_{\F}^2  \nonumber\\
    &=D - \,\text{tr}\big((V^{\T}V)^{-1}V^{\T}Y(Y^{\T}Y)^{-1}Y^{\T}V\big)
\end{align}

To learn the embeddings, bi-directional LSTM (BLSTM) is usually used to model the context information from past and future frames. The network architecture is shown at the bottom of Fig.~\ref{fig:MISI}, where the DC embedding layer is a fully-connected layer with a non-linearity such as a logistic sigmoid, followed by unit-length normalization for each frequency. 

Another permutation-free training scheme was proposed for mask-inference networks first in \cite{Hershey2016}, and was later found to be working very well in \cite{Isik2016} and \cite{Yu2017a}. The idea is to train a mask-inference network to minimize the minimum loss over all permutations. Following \cite{Kolbæk2017}, the phase-sensitive mask (PSM) \cite{Erdogan2015} is used as the training target. It is common in phase-sensitive spectrum approximation (PSA) to truncate the unbounded mask values. Using   $\operatorname{T}_{a}^{b}(x)= \min(\max(x,a),b)$, the truncated PSA (tPSA) objective is
\begin{multline}
	\mathcal{L}_{\text{tPSA}} = \min_{\pi \in \mathcal{P}} \sum_{c} \Big\| \hat{M}_{\pi(c)} \odot |X| 
	\\ - \operatorname{T}_{0}^{\gamma|X|}\left(|S_c| \odot \cos(\angle S_c - \angle X)\right) \Big\|_1, \label{eq:L_MI_PSA}
\end{multline}
where $\angle X$ is the mixture phase, $\angle S_c$ the phase of the $c$-th source, $\mathcal{P}$ the set of permutations on $\{1,\dots,C\}$, $|X|$ the mixture magnitude, $\hat{M}_c$ the $c$-th estimated mask, $|S_c|$ the magnitude of the $c$-th reference source, $\odot$ denotes element-wise matrix multiplication, and $\gamma$ is a mask truncation factor. Sigmoidal activation together with $\gamma=1$ is commonly used in the output layer of T-F masking. To endow the network with more capability, multiple activation functions that can work with $\gamma>1$ will be discussed in Section \ref{activation}. 

Following \cite{Luo2017}, our recent study \cite{Wang2018ICASSP} proposed a chimera++ network combining the two approaches via multi-task learning, as illustrated in the bottom of Fig.~\ref{fig:MISI}.
The loss function is a weighted sum of the deep clustering loss and the mask inference loss.
\begin{align} \label{chimera:1}
	\mathcal{L}_{\text{chi}^{++}_\alpha}=\alpha \mathcal{L}_{\text{DC,W}}+(1-\alpha)\mathcal{L}_{\text{tPSA}}
\end{align}
Only the MI output is needed to make predictions at run time.

\section{Proposed Algorithms}
\subsection{Iterative Phase Reconstruction}
There are multiple target sources to be separated in each mixture in our study. The Griffin-Lim algorithm \cite{Griffin1984a} only performs iterative reconstruction for each source independently. In \cite{Wang2018ICASSP}, we therefore proposed to utilize the MISI algorithm \cite{Gunawan2010} (see Algorithm \ref{MISI}) to reconstruct the clean phase of each source starting from the estimated magnitude of each source and the mixture phase, where the sum of the reconstructed time-domain signals after each iteration is constrained to be the same as the mixture signal. Note that the estimated magnitudes remain fixed during iterations, while the phase of each source are iteratively reconstructed. In \cite{Wang2018ICASSP}, the phase reconstruction was only added as a post-processing, and it was not part of the objective function during training, which remained computed on the time-frequency representation of the estimated signal, prior to resynthesis. In this paper, we go several steps further.

\begin{algorithm}[!tp] 
\SetKwInOut{Input}{Input}\SetKwInOut{Output}{Output}
{\footnotesize 
	\caption{\small Iterative phase reconstruction based on MISI. STFT($\cdot$) extracts the STFT magnitude and phase of a signal, and iSTFT($\cdot$,$\cdot$) reconstructs a time-domain signal from a magnitude and a phase.\vspace{-0.5cm}} \label{MISI}
	\Input{Mixture time-domain signal $x$, mixture complex spectrogram $X$, mixture phase $\angle X$, enhanced magnitudes ${\hat{A}}_c=\hat{M}_c \circ |X|$ for $c=1,\dots,C$, and iteration number $K$}
	\Output{Reconstructed phase $\hat{\theta}_{c}^{(K)}$ and signal $\hat{s}_{c}^{(K)}$ for $c=1,\dots,C$}
	$\hat{s}_{c}^{(0)}=\text{iSTFT}({\hat{A}}_c,\angle X)$, for $c=1,\dots,C$\;
	\For{i = $1,\dots,K$ }{
		$\delta^{(i-1)}=x-\sum_{c=1}^C \hat{s}_{c}^{(i-1)}$\;
		$\hat{\theta}_{c}^{(i)}= \angle \text{STFT}\big(\hat{s}_{c}^{(i-1)}+\frac{\delta^{(i-1)}}{C}\big)$, for $c=1,\dots,C$\;
		$\hat{s}_{c}^{(i)}=\text{iSTFT}({\hat{A}}_c,\hat{\theta}_{c}^{(i)})$, for $c=1,\dots,C$\;
	}
}
\end{algorithm}\vspace{-0.25cm}

\subsection{Waveform Approximation}
The first step in phase reconstruction algorithms such as MISI is to reconstruct a waveform from a time-frequency domain representation using the inverse STFT.
We thus consider a first objective function computed on the waveform reconstructed by iSTFT, denoted as waveform approximation (WA), and represent iSTFT as various layers on top of the mask inference layer, so that end-to-end optimization can be performed. The label permutation problem is resolved by minimizing the minimum $L_1$ loss of all the permutations at the waveform level. We denote the model trained this way as WA. The objective function to train this model is
\begin{align} 
	\mathcal{L}_{\text{WA}} = \min_{\pi \in \mathcal{P}} \sum_{c} \Big\| \hat{s}_{\pi(c)}^{(0)} - s_{c} \Big\|_1, \label{eq:L_WA}
\end{align}
where $s_{c}$ denotes the time-domain signal of source $c$, and $\hat{s}_c^{(0)}$ denotes the $c$-th time-domain signal obtained by inverse STFT from the combination of the $c$-th estimated magnitude and the mixture phase. Note that mixture phase is still used here and no phase reconstruction is yet performed. This corresponds to the initialization step in Algorithm~\ref{MISI}.

In \cite{wang2015}, a time-domain reconstruction approach is proposed for speech enhancement. However, their approach only trains a feed-forward mask-inference DNN through iDFT separately for each frame using squared error in the time domain. By Parseval's theorem, this is equivalent to optimizing the mask for minimum squared error in the complex spectrum domain, when using the noisy phases, as in \cite{Erdogan2015}, proposed in the same conference. A follow-up work \cite{Zhao2017b} of \cite{wang2015} supplies clean phase during training. However, this makes their approach equivalent to conventional magnitude spectrum approximation \cite{Weninger2014}, which does not perform as well as the phase-sensitive mask \cite{Weninger2015}. Closest to the above WA objective, an adaptive front-end framework was recently proposed \cite{venkataramani2017end} in which the STFT and its inverse are subsumed by the network, along with the noisy phase, so that training is effectively end-to-end in the time-domain. The proposed method then replaces the STFT and its inverse by trainable linear convolutional layers.  Unfortunately the paper does not compare training through the STFT to the conventional method so the results are uninformative about this direction.  

\subsection{Unfolded Iterative Phase Reconstruction}
\begin{figure}  
	\centering  
	\includegraphics[width=8cm]{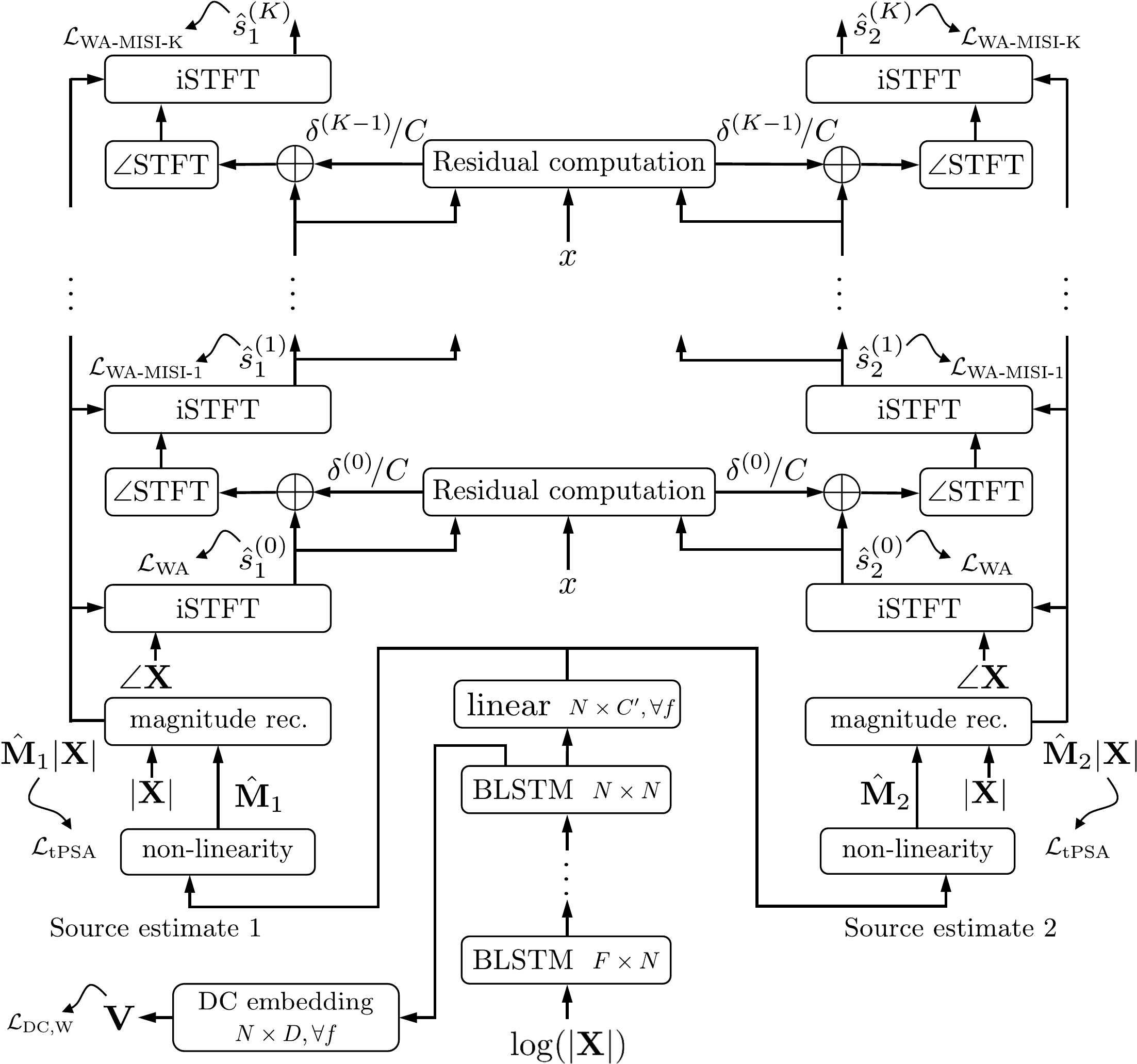} \vspace{-0.3cm}
	\caption{Training through $K$ MISI iterations.}
	\label{fig:MISI}  \vspace{-0.4cm}
\end{figure}

We further unfold the iterations in the MISI algorithm as various deterministic layers in a neural network. This can be achieved by further growing several layers representing STFT and iSTFT operations on top of the mask inference layer. By performing end-to-end optimization that trains through MISI, the network can become aware of the later iterative phase reconstruction steps and learn to produce estimated magnitudes that are well-suited to that subsequent processing, hence producing better phase estimates for separation. The model trained this way is denoted as WA-MISI-K, where $K\geq 1$ is the number of unfolded MISI iterations. The objective function is
\begin {align}
	\mathcal{L}_{\text{WA-MISI-K}} = \min_{\pi \in \mathcal{P}} \sum_{c} \Big\| \hat{s}_{\pi(c)}^{(K)} - s_{c} \Big\|_1, \label{eq:L_WA_MISI}
\end {align}
where $\hat{s}_c^{(K)}$ denotes the $c$-th time-domain signal obtained after $K$ MISI iterations as described in Algorithm \ref{MISI}. The whole separation network, including unfolded phase reconstruction steps at the output of the mask inference head of the Chimera++ network, is illustrated in Fig.~\ref{fig:MISI}.
The STFT and iSTFT can be easily implemented using modern deep learning toolkits as deterministic layers efficiently computed on a GPU and through which backpropagation can be performed.

A recent study by Williamson et al.\ \cite{Williamson2016, Williamson2017a} proposed a complex ratio masking approach for phase reconstruction and speech enhancement, where a feed-forward DNN is trained to predict the real and imaginary components of the ideal complex filter in the STFT domain, i.e., $M_c=S_c/X=|S_c|e^{j(\angle S_c-\angle X)}/|X|$ for source $c$ for example. The real component is equivalent to the earlier proposed phase-sensitive mask \cite{Erdogan2015}, which contains patterns clearly predictable from energy-based features \cite{Erdogan2015, Weninger2015}. However, recent studies along this line suggest that the patterns in the imaginary component are too random to predict \cite{Williamson2017}, possibly because it is difficult for a learning machine to determine the sign of $\sin(\angle S_c-\angle X)$ only from energy-based features. In contrast, the $\cos(\angle S_c-\angle X)$ in the real component is typically much smaller than one for T-F units dominated by other sources and close to one otherwise, making itself predictable from energy-based features. The proposed method thus only focuses on estimating a mask in the magnitude domain and uses the estimated magnitude to elicit better phase through  iterative phase reconstruction.

Another recent trend is to avoid the phase inconsistency problem altogether by operating in the time domain, using convolutional neural networks \cite{Fu2017a, Fu2017b}, WaveNet \cite{Qian2017}, generative adversarial networks \cite{Pascual2017}, or encoder-decoder architectures \cite{Luo2017a}. Although they are promising approaches, the current state-of-the-art approach for supervised speech separation is via T-F masking \cite{Wang;, Wang2018ICASSP}. The proposed approach is expected to produce even better separation if the phase can be reconstructed.

\subsection{Activation Functions with Values Beyond One} \label{activation}
Sigmoidal units are dominantly used in the output layer of deep learning based T-F masking \cite{Wang2017b, Wang;}, partly because they can model well data with bi-modal distribution \cite{Bishop2006}, such as the IRM \cite{Wang2014} and its variants \cite{Wang2017b}. Restricting the possible values of the T-F mask to lie in $[0,1]$ is also reasonable when using the mixture phase for reconstruction: indeed, T-F mask values larger than one would in theory be needed in regions where interferences between sources result in a mixture magnitude smaller than that of a source; but the mixture phase is also likely to be different from the phase of that source in such regions, in which case it is more rewarding in terms of objective measure to oversuppress than to go even further in a wrong direction. This is no longer valid if we consider phase reconstruction in the optimization. Moreover, capping the mask values to be between zero and one is more likely to take the enhanced magnitude further away from the consistent STFT domain, posing potential difficulties for later phase reconstruction.

To obtain clean magnitudes, the oracle mask should be $|S_c|/|X|$ 
(also known as the FFT mask in \cite{Wang2014} or the ideal amplitude mask in \cite{Erdogan2015}). Clearly, this mask can go beyond one, because the underlying sources, although statistically independent, may have opposite phase at a particular T-F unit, therefore cancelling with each other and producing a mixture magnitude that is smaller than the magnitude of a given source. It is likely much harder to predict the mask values of such T-F units, but we believe that it is still possible based on contextual information.

In our study, we truncate the values in PSM to the range $[0,2]$ (i.e., $\gamma=2$ in Eq.~(\ref{eq:L_MI_PSA})), as only a small percentage of mask values goes beyond this range. Multiple activation functions can be utilized in the output layer. We here consider:
\begin{itemize}[leftmargin=4mm, topsep=0pt]
    \setlength{\itemsep}{0pt}
    \setlength{\parsep}{0pt}
    \setlength{\parskip}{0pt}
    \item doubled sigmoid: sigmoid non-linearity multiplied by 2;
    \item clipped ReLU: ReLU non-linearity clipped to $[0,2]$;
    \item convex softmax: the output non-linearity is a three-dimensional softmax for each source at each T-F unit. It is used to compute a convex sum between the values 0, 1, and 2: $y= [x_0,x_1,x_2] [0,1,2]^T$ where $[x_0,x_1,x_2]$ is the output of the softmax. This activation function is designed to model the three modes concentrated at 0, 1 and 2 in the histogram of the PSM.
\end{itemize}

\begin{table}[t]

    \footnotesize
	\centering
	\caption{SI-SDR (dB) performance on wsj0-2mix.} \vspace{-0.35cm}
	\label{table 1}
	\begin{tabular}{lcc}
		\hline\hline
        Approaches & CSC & OSC   \\ 
		\hline\hline
		$\mathcal{L}_{\text{DC,W}}$ & 10.4 & 10.4   \\
		\hline
		$\mathcal{L}_{\text{tPSA}}$ & 10.1 & 10.0 \\ 
		\hline
		$\mathcal{L}_{\text{chi}^{++}_\alpha}$ (sigmoid) & 11.1 & 11.2 \\
		\quad + Griffin-Lim-5 & 11.2 &11.3 \\
		\quad + MISI-5 & 11.4 &11.5 \\
		\quad + $\mathcal{L}_{\text{WA}}$ & 11.6 & 11.6 \\ 
		\quad\quad +MISI-5 & 11.6 & 11.6 \\ 
		\quad\quad +$\mathcal{L}_{\text{WA-MISI-5}}$ & 12.4 & 12.2 \\ 
	    \hline
		$\mathcal{L}_{\text{chi}^{++}_\alpha}$ (doubled sigmoid) & 10.0 & 10.0 \\ 
		\quad +$\mathcal{L}_{\text{WA}}$ & 11.5 & 11.4 \\ 
		\quad\quad +$\mathcal{L}_{\text{WA-MISI-5}}$ & 12.5 & 12.3 \\ 
	    \hline
		$\mathcal{L}_{\text{chi}^{++}_\alpha}$(clipped RelU) & 10.4 & 10.4 \\ 
		\quad +$\mathcal{L}_{\text{WA}}$ & 11.7 & 11.7 \\ 
		\quad\quad +$\mathcal{L}_{\text{WA-MISI-5}}$ & 12.6 & 12.4 \\ 
	    \hline
		$\mathcal{L}_{\text{chi}^{++}_\alpha}$(convex softmax) & 11.0 & 11.1 \\ 
		\quad +$\mathcal{L}_{\text{WA}}$ & 11.8 & 11.8 \\ 
		\quad\quad +$\mathcal{L}_{\text{WA-MISI-5}}$ & \textbf{12.8} & \textbf{12.6} \\ 
		\hline\hline
	\end{tabular}\vspace{-0.2cm}
\end{table}

\section{Experimental Setup}
\label{sec:setup}
We validate the proposed algorithms on the publicly-available wsj0-2mix corpus \cite{Hershey2016}, which is widely used in many speaker-independent speech separation tasks. It contains 20,000, 5,000 and 3,000 two-speaker mixtures in its 30~h training, 10~h validation, and 5~h test sets, respectively. The speakers in the validation set (closed speaker condition, CSC) are seen during training, while the speakers in the test set (open speaker condition, OSC) are completely unseen. The sampling rate is 8~kHz.

Our neural network contains four BLSTM layers, each with 600 units in each direction. A dropout of $0.3$ is applied on the output of each BLSTM layer except the last one. The network is trained on 400-frame segments using the Adam algorithm. The window length is 32~ms and the hop size is 8~ms. The square root Hann window is employed as the analysis window and the synthesis window is designed accordingly to achieve perfect reconstruction after overlap-add. A 256-point DFT is performed to extract 129-dimensional log magnitude input features. We first train the chimera++ network with $\alpha$ set to 0.975. Next, we discard the deep clustering branch (i.e., we set $\alpha$ to 0) and train the network with $\mathcal{L}_{\text{WA}}$. Subsequently, the network is trained using $\mathcal{L}_{\text{WA-MISI-1}}$, then $\mathcal{L}_{\text{WA-MISI-2}}$, and all the way to $\mathcal{L}_{\text{WA-MISI-K}}$, where here $K=5$, as performance saturated after five iterations in our experiments. We found this curriculum learning strategy to be helpful. At run time, for the models trained using $\mathcal{L}_{\text{WA-MISI-5}}$, we run MISI with $5$ iterations, while results for other models are obtained without phase reconstruction unless specified. 

We report the performance using \emph{scale-invariant SDR} (SI-SDR) \cite{Hershey2016,Isik2016,Luo2018,LeRoux2018SISDR}, as well as the SDR metric computed using the bss\_eval\_sources software \cite{Vincent2006BSSeval} because it is used by other groups. We believe SI-SDR is a more proper measure for single-channel instantaneous mixtures \cite{LeRoux2018SISDR}.

\begin{figure}[t]
	\centering  
	\includegraphics[width=8cm]{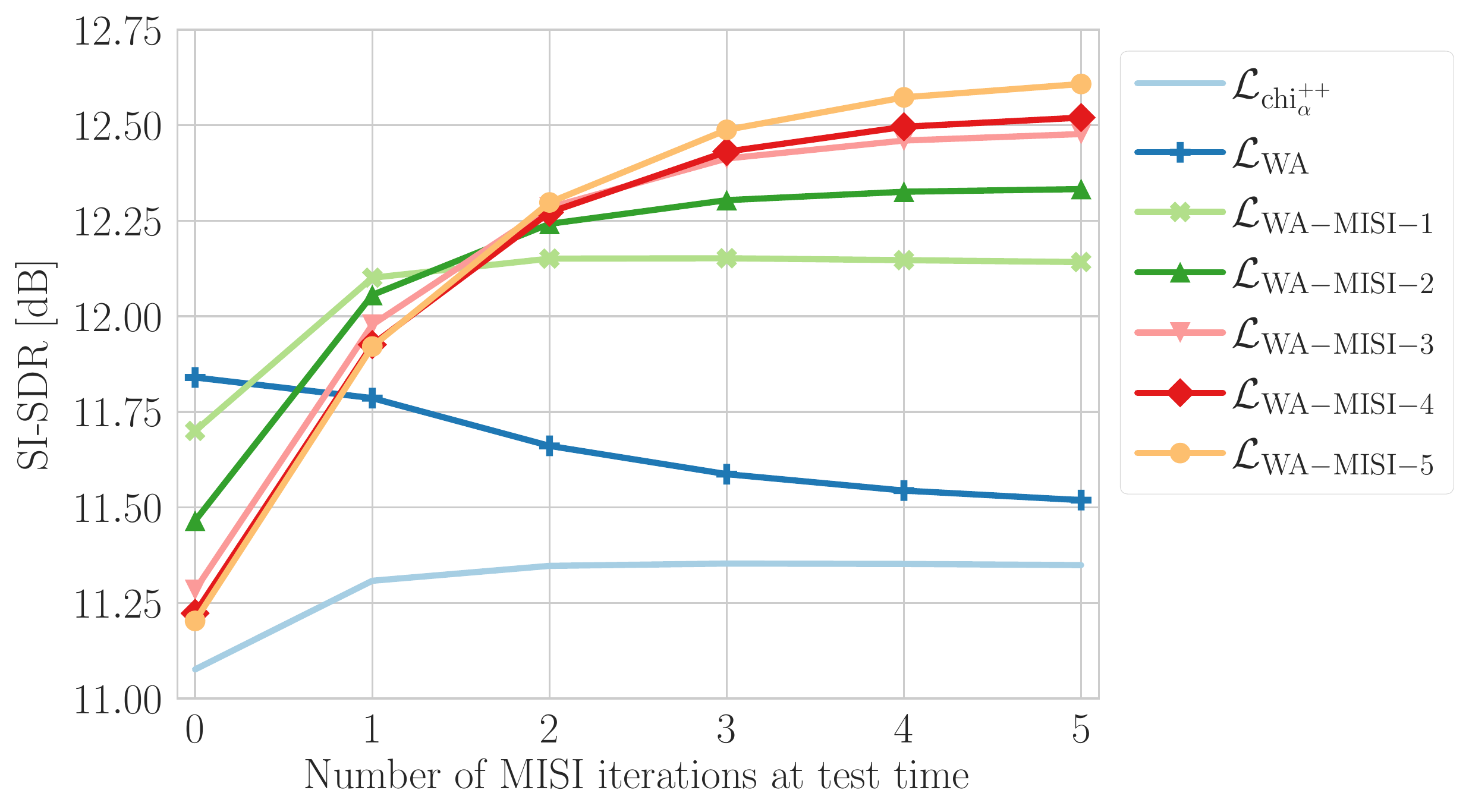} \vspace{-0.4cm}
	\caption{SI-SDR vs number of MISI iterations at test time}
	\label{fig:SDR_vs_MISI_iterations}  \vspace{-0.4cm}
\end{figure}

\begin{table}[t]

    \footnotesize
	\centering
	\caption{Comparison with other systems on wsj0-2mix.} \vspace{-0.35cm}
	\label{table 2}
	\begin{tabular}{l|cc|cc}
		\hline\hline
		\multicolumn{1}{c|}{} & \multicolumn{2}{c|}{SI-SDR (dB)} & \multicolumn{2}{c}{SDR (dB)} \\ \cline{2-5} 
\multicolumn{1}{c|}{Approaches}  & CSC & OSC & CSC & OSC \\
		\hline\hline
		Deep Clustering \cite{Hershey2016, Isik2016} & - & 10.8 & - & - \\
		\hline
		Deep Attractor Networks \cite{Chen2017, Luo2018}  & - & 10.4 & - & 10.8 \\
		\hline
		PIT \cite{Yu2017a, Kolbæk2017} & - & - & 10.0 & 10.0 \\
		\hline
        TasNet \cite{Luo2017a} & - & 10.2 & - & 10.5 \\
		\hline
        Chimera++ Networks \cite{Wang2018ICASSP} & 11.1 & 11.2 & 11.6 & 11.7 \\
        \quad + MISI-5 \cite{Wang2018ICASSP} & 11.4 & 11.5 & 12.0 & 12.0 \\
        \hline
        WA (proposed) & 11.8 & 11.8 & 12.3 & 12.3 \\
        WA-MISI-5 (proposed) & \textbf{12.8} & \textbf{12.6} & \textbf{13.2} & \textbf{13.1}\\
		\hline
		Oracle Masks: & & \\
		\quad Magnitude Ratio Mask & 12.5 & 12.7 & 13.0 & 13.2 \\
		\quad \quad + MISI-5 & 13.5 & 13.7 & 14.1 & 14.3\\
		\quad Ideal Binary Mask & 13.2 & 13.5 & 13.7 & 14.0 \\ 
		\quad \quad + MISI-5 & 13.1 & 13.4 & 13.6 & 13.8 \\
		\quad PSM & 16.2 & 16.4 & 16.7 & 16.9 \\ 
		\quad \quad + MISI-5 & 18.1 & 18.3 & 18.5 & 18.8 \\
		\quad Ideal Amplitude Mask & 12.6 & 12.8 & 12.9 & 13.2 \\ 
		\quad \quad + MISI-5 & 26.3 & 26.6 & 26.8 & 27.1 \\
		\hline\hline
	\end{tabular}\vspace{-0.4cm}
\end{table}

\section{Evaluation Results}
\label{sec:exp}

Table \ref{table 1} reports the SI-SDR results on the wsj0-2mix dataset. We first present the results using sigmoidal activation. The chimera++ network obtains significantly better results than the individual models (11.2 dB vs.\ 10.4 dB and 10.0 dB SI-SDR). With the mixture phase and estimated magnitudes, performing five iterations of MISI pushes the performance to 11.5 dB, while 11.3 dB is obtained when applying five iterations of Griffin-Lim on each source independently, as is reported in \cite{Wang2018ICASSP}. Performing end-to-end optimization using  $\mathcal{L}_{\text{WA}}$ improves the results to 11.6 dB from 11.2 dB, without requiring phase reconstruction post-processing. Further applying MISI post-processing for five iterations (MISI-5) on this model however does not lead to any improvements, likely because the mixture phase is used during training and the model compensates for it without expecting further processing. In contrast, training the network through MISI using $\mathcal{L}_{\text{WA-MISI-5}}$ pushes the performance to 12.2 dB. 

Among the three proposed activation functions, the convex softmax performs the best, reaching 12.6 dB SI-SDR. It thus seems effective to model the multiple peaks in the histogram of the truncated PSM, and important to produce estimated magnitudes that are closer to the consistent STFT domain. As expected, activations going beyond $1$ only become beneficial when training through phase reconstruction.

In Fig.~\ref{fig:SDR_vs_MISI_iterations}, we show the evolution of the SI-SDR performance of the convex softmax models trained with different objective functions against the number of MISI iterations at test time ($0$ to $5$). Training with $\mathcal{L}_{\text{WA}}$ leads to a magnitude that is very well suited to iSTFT, but not to further MISI iterations. As we train for more MISI iterations, performance starts lower, but reaches higher values with more test-time iterations.

Table \ref{table 2} lists the performance of competitive approaches on the same corpus, along with the performance of various oracle masks with or without applying MISI for five iterations. The first three algorithms use mixture phase directly for separation. The fourth one, time-domain audio separation network (TasNet), operates directly in the time domain. Our result is 1.1 dB better than the previous state-of-the-art by \cite{Wang2018ICASSP} in terms of both SI-SDR and SDR.

\section{Concluding Remarks}
\label{sec:conclusion}
We have proposed a novel end-to-end approach for single-channel speech separation. Significant improvements are obtained by training the T-F masking network through an iterative phase reconstruction procedure. Future work includes applying the proposed methods to speech enhancement, considering the joint estimation of magnitude and an initial phase that improves upon the mixture phase, and improving the estimation of the ideal amplitude mask. We shall also consider alternatives to the waveform-level loss, such as errors computed on the magnitude spectrograms of the reconstructed signals. 

\vfill\pagebreak
\balance

\bibliographystyle{IEEEtran_nourl}
\bibliography{unfolded_phase}

\end{document}